\documentclass{elsart}
\usepackage{graphics,graphicx,amssymb}
\newbox\grsign \setbox\grsign=\hbox{$>$} \newdimen\grdimen \grdimen=\ht\grsign
\newbox\simlessbox \newbox\simgreatbox
\setbox\simgreatbox=\hbox{\raise.5ex\hbox{$>$}\llap
     {\lower.5ex\hbox{$\sim$}}}\ht1=\grdimen\dp1=0pt
\setbox\simlessbox=\hbox{\raise.5ex\hbox{$<$}\llap
     {\lower.5ex\hbox{$\sim$}}}\ht2=\grdimen\dp2=0pt

\begin{document}

\begin{frontmatter}

\title{An almost linear stochastic map related to the particle system models of social sciences}

\author{Anindya S. Chakrabarti}

\textit{\small{Economics Department, Boston University, 270 Bay State Road, MA- 02215, USA}}


\thanks{Email addresses: aschakrabarti@gmail.com}

\begin{abstract}
\noindent  We propose a stochastic map model of economic dynamics. 
In the last decade, an array of observations in economics has been investigated in the econophysics literature, 
a major example being the universal
features of inequality in terms of income and wealth. 
Another area of inquiry is the formation of opinion in a society. 
The proposed model attempts to produce positively skewed distributions and the power law distributions
as has been observed in the real data of income and wealth. Also, it shows
a non-trivial phase transition in the opinion of a society (opinion formation). A number of physical models
also generates similar results.
In particular, 
the kinetic exchange models have been especially successful in this regard. Therefore, we compare the
results obtained from these two approaches and discuss a number of new features and drawbacks of this model.

\end{abstract}
\end{frontmatter}

\section{Introduction}

\noindent It is known that the distributions of income
and wealth possess some robust and stable
features which are independent of the economy-specific conditions \cite{yako-rosser;09}. But
the exact form of the distribution is still debated \cite{kleiber;02}.
It has been a tradition in the economics literature to model the left tail and the mode of the distributions of 
the incomes
with a log-normal \cite{gibrat;31}
distribution and the right tail with a Pareto distribution i.e., a power law \cite{pareto;97}. 
However, a number of recent studies in econophysics literature argue that
the left tail and the mode of
the distribution is best described by the exponential or gamma distribution 
and the right tail of the distribution follows a power
law (See Ref. \cite{drag-yako;01, drag-yako;01a, acbkc;07}). 
Also, a significant number of attempts have been devoted to explain the emergence of a consensus in a society
and the possibility of a non-trivial phase-transition in the average opinion of the society 
(See Ref. \cite{weisbuch,lallouache;10}).

\noindent Here, we propose a stochastic map which attempts to produce all of the aforementioned
features at different limits and in addition, it shows some new features as well. Below, we propose the model and discuss
its relative merits and demerits with respect to the kinetic exchange models which also generates similar results. For
related literature, see Ref. \cite{yako-rosser;09, kwem, acbkc;07,chakrabartis;10} 
for theoretical and numerical results on the kinetic exchange models of markets. 
See Ref. \cite{kesten, vervaat, babillot} for detailed analysis of the limit distributions of the solution
of random difference equations (of the form $x(t)=a(t)x(t-1)+b(t)$; 
we will use only a few very particular instances of it). Ref. \cite{tollar} characterizes gamma distribution
which arises from a random difference equation.
Ref. \cite{mohanty} was the first paper that suggested that multi-agent interactions in the kinetic wealth 
exchange models
can be simplified and
and viewed from from the point of view of a single agent. Ref. \cite{sinha;03} was possibly the first paper
that connected the stochastic maps and the kinetic exchange models.

\section{The map}
\label{sec:map}

\noindent The Kolkata kinetic wealth exchange models have been successful to generate a realistic description
of the income/wealth distributions (See ref. \cite{acbkc;07}). Many-body interactions are the key
ingredients of this class of (kinetic) exchange models. In this paper, we mainly use a single agent
framework to discuss the distributional issues. There is no agent-agent interaction in this model. In this regard, our treatment is closer to the
{\it representative} agent paradigm of the modern economics. The specific structure of the market  
is also not discussed here. We represent the interaction of the agent with the market with a map of the following form, 

\begin{equation}
m(t) = min~\{(\lambda_1+\epsilon_t \lambda_2) m(t-1) +\xi_t \lambda^{n}_3, \theta\} 
\label{map}
\end{equation}

\noindent where $\lambda_1$, $\lambda_2$ and $\lambda_3$ are linear functions of a single parameter $\lambda$ with 
$0\le\lambda\le 1$ such that $0\le \lambda_i\le 1$ for $i$=1, 2 and 3. 
We assume that $-\infty \le n \leq 1$, $\theta$ takes value either a positive, finite value (we assume $\theta$ = 1,
for convinience) 
or $\infty$ (or a sufficiently large value) and 
$\epsilon_t, \xi_t \sim uniform[0,1]$ and independent, unless specified. In all the simulations, we
have assumed $m(0)$ = 1.
We denote the time index by subscripts for exogenous random variables ($\epsilon$ and $\xi$) which we shall drop
when no confusion arises.
While the parameter space is seemingly too large to be considered in details, we shall 
restrict it considerably by assuming very simple forms of $\lambda_i$ for all $i$.

\noindent We can interpret it in the following way. An agent has $m(t)$ amount of wealth (or money) at time $t$ of which
he saves a random fraction and invests the rest. His return from investment is represented by the additive term 
$\xi_t \lambda^{n}_3$. However, there is an upper limit of $m(t)$ represented by $\theta$.
Since we interpret
the random multiplicative term as the savings propensity, we assume that $\lambda_1+\epsilon_t \lambda_2 \le 1$ for all $t$
i.e., savings propensity is never greater than 1. Our interest lies in finding the pdf (probability density function
$P(m)$) in the steady state.

\noindent It may be noted that the above map (ignoring $\theta$) has the general form

\begin{equation}
m(t)= a(t) m(t-1)+ b(t)
\label{kesteneqn}
\end{equation}

\noindent which has been studied in great details by Ref. \cite{kesten,vervaat,babillot}. In this paper, we have focused
on a few particular instances of it for our purpose. Ref. \cite{sinha;03} mapped the asset exchange models
into `random iterated function systems'. But it was concerned with the `yard-sale' model and the `theft-and-fraud'
model (see Ref. \cite{hayes;02}) whereas we focus on the CC and CCM models (the models that introduced
fixed and distributed savings propensities in the kinetic exchange models; see Ref. \cite{acbkc;07}) and that is
why we have borrowed the basic structure (a fraction of the total wealth is 
saved and the rest is invested) from the Kolkata wealth exchange models (i.e., the CC and CCM models).  

\medskip

\section{Statistical features of the map at some limits}

\subsection{A positively skewed distribution}
\label{subsec: Case 1}
\noindent 

\noindent We assume $\lambda_1 = \lambda$, $\lambda_2 = \lambda_3 = 1-\lambda$, $n = 1 $ and $\theta \rightarrow \infty$.
The equation becomes

\begin{equation}
m(t+1)= \{\lambda+\epsilon(1-\lambda)\}m(t)+\xi(1-\lambda).
\label{gamma1}
\end{equation}

\begin{figure}
\begin{center}
\noindent \includegraphics[clip,width= 6cm, angle = 270]
{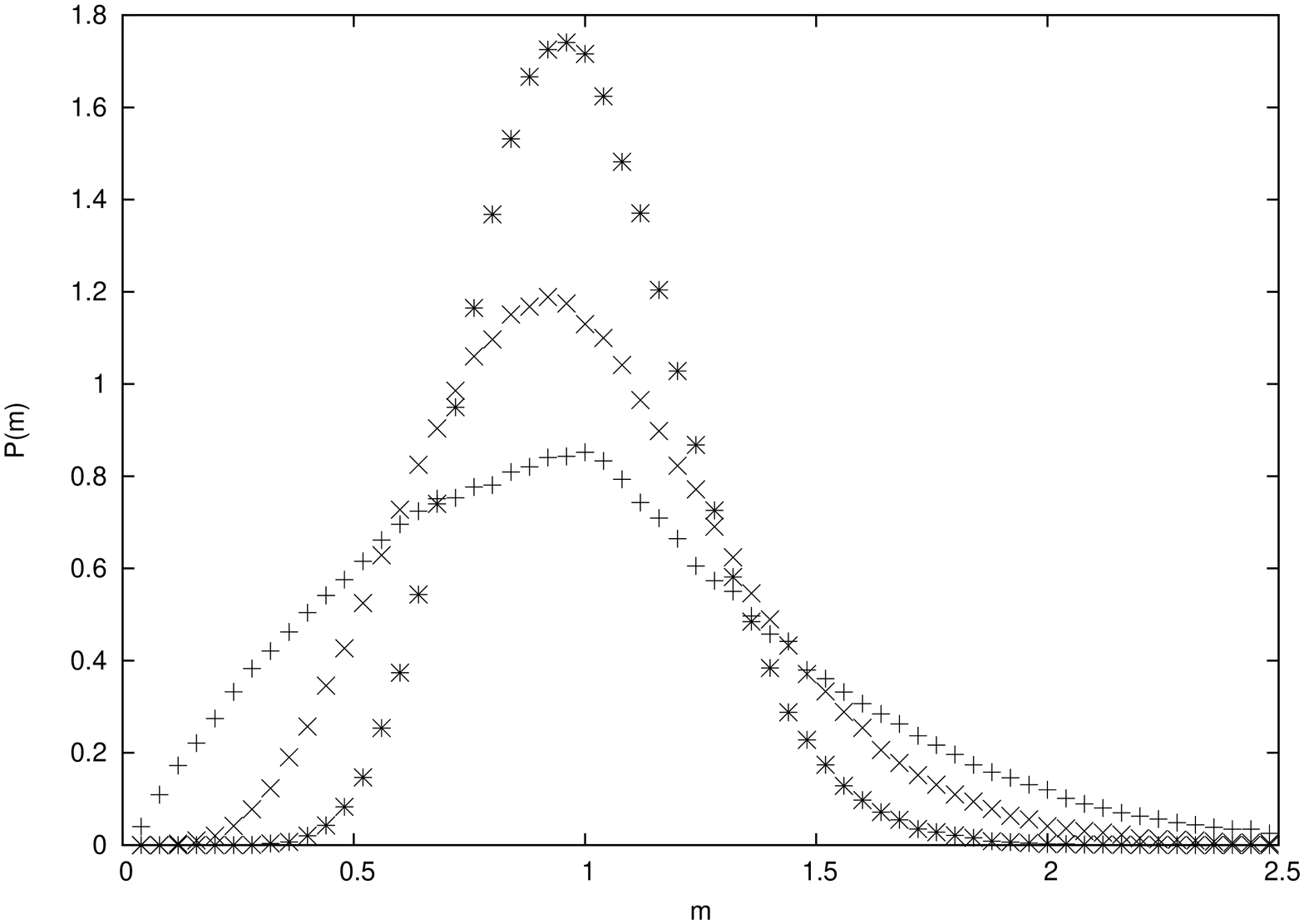}
\caption\protect{Steady state distributions generated by Eqn. \ref{gamma1}: Three cases are shown
above, viz., $\lambda=0$ (+), $\lambda=0.4$ ($\times$),
$\lambda=0.7$ ($\ *$). All simulations are done for $\sim ~10^5$ time steps and averaged over for $\sim ~10^4$. 
}
\label{almostgamma}
\end{center}
\end{figure}

\noindent See Fig. \ref{almostgamma} for steady state distributions corresponding to different values of $\lambda$. 
We restrict our attention to the case
$n~=~1$ as opposed to other positive numbers, for two reasons. 
One, this choice of $n$ makes the average a constant, independent of $n$ (as the CC and CCM models; 
see Ref. \cite{acbkc;07}) and secondly, to simplify the calculations. We will relax
this assumption in Section \ref{subsec: Case 2} where we will assume $n$ is negative and we will see that
the exponents of the power law distributions differ for different values of $n$.  

Though we do not know the exact 
algebraic form of the steady state distributions produced by Eqn. \ref{gamma1}, we can
find out its moments in order to describe them qualitatively. 
We can ignore the time index in the steady state.
Taking expectations over both sides of Eqn. \ref{gamma1}, we get
\begin{equation}
\langle m \rangle =1.
\end{equation}
\noindent

\noindent Also, we have the variance of $m$ as
\begin{equation}
V(m)=\langle x^2\rangle-\langle x\rangle^2
\label{variance}
\end{equation} 

\noindent where $x=\left(\lambda+\epsilon(1-\lambda)\right)m+\xi(1-\lambda)$.
Now, we make a few almost trivial observations.
Note the fact that $\langle x\rangle = 1$. Hence, $\langle m^2\rangle$ can be written as 
$V(m)+\langle m\rangle^2$ (by defn. of $V(m)$) i.e., $V(m)+1$. 
Also, $\epsilon$ and $\xi$ are uniformly distributed. Therefore, $\langle \epsilon \rangle$ = 1/2 = 
$\langle \xi \rangle$
and $\langle \epsilon^2 \rangle$ = 1/3 = $\langle \xi^2 \rangle$ (recall that $V(\epsilon)$ = 1/12 = $V(\xi)$).
Using all of these and by expanding Eqn. \ref{variance} we get

\begin{eqnarray}
V(m) &=& [\lambda^2+\lambda(1-\lambda)+\frac{1}{3}(1-\lambda)^2] \left(V(m)+1\right) \nonumber\\
&& +\frac{1}{3}(1-\lambda)^2+\frac{1}{2}(1-\lambda^2)-1. \nonumber
\end{eqnarray}

\noindent Simplifying the above expression we get the result for $\lambda\ne 1$,
\begin{equation}
V(m)=\frac{1}{2}\left(\frac{1-\lambda}{2+\lambda}\right)
\label{gamma2}
\end{equation}
\noindent which clearly shows that the distribution tends to a delta function as $\lambda\rightarrow 1$. 
With $\lambda$ = 0, Eqn. \ref{gamma1} produces 
a distribution with a sharp peak at $m$ very close to 1. However, for $\lambda~>~0.3$ this sharpness goes away.

\noindent One can consider an even simpler case with $\xi_t$ =$\epsilon_t$ for all $t$ so that the map becomes  

\begin{equation}
m(t+1)= \lambda m(t)+\epsilon(1-\lambda)\left(m(t)+1\right)
\label{gamma3}
\end{equation}

\noindent which is almost the same as the usual kinetic exchange model with a constant savings
factor i.e., the CC model (See Ref. \cite{chakraborti;00}; See also Ref. \cite{acbkc;07}),
\begin{equation}
m_i(t+1)= \lambda m_i(t)+\epsilon(1-\lambda)\left(m_i(t)+m_j(t)\right).
\label{gamma4}
\end{equation}

\noindent The variance of the distribution generated by Eqn.\ref{gamma3} is given by

\begin{equation}
V(m)=\left(\frac{1-\lambda}{2+\lambda}\right).
\label{gamma5}
\end{equation}

\noindent For the sake of completeness, we mention that the variance of the distribution generated by Eqn. \ref{gamma4}
is given by (See Ref. \cite{richmond;05})

\begin{equation}
V(m)=\frac{(1-\lambda)}{(1+2\lambda)}.
\label{gamma6}
\end{equation}

\begin{figure}
\begin{center}
\noindent \includegraphics[clip,width= 6cm, angle = 270]
{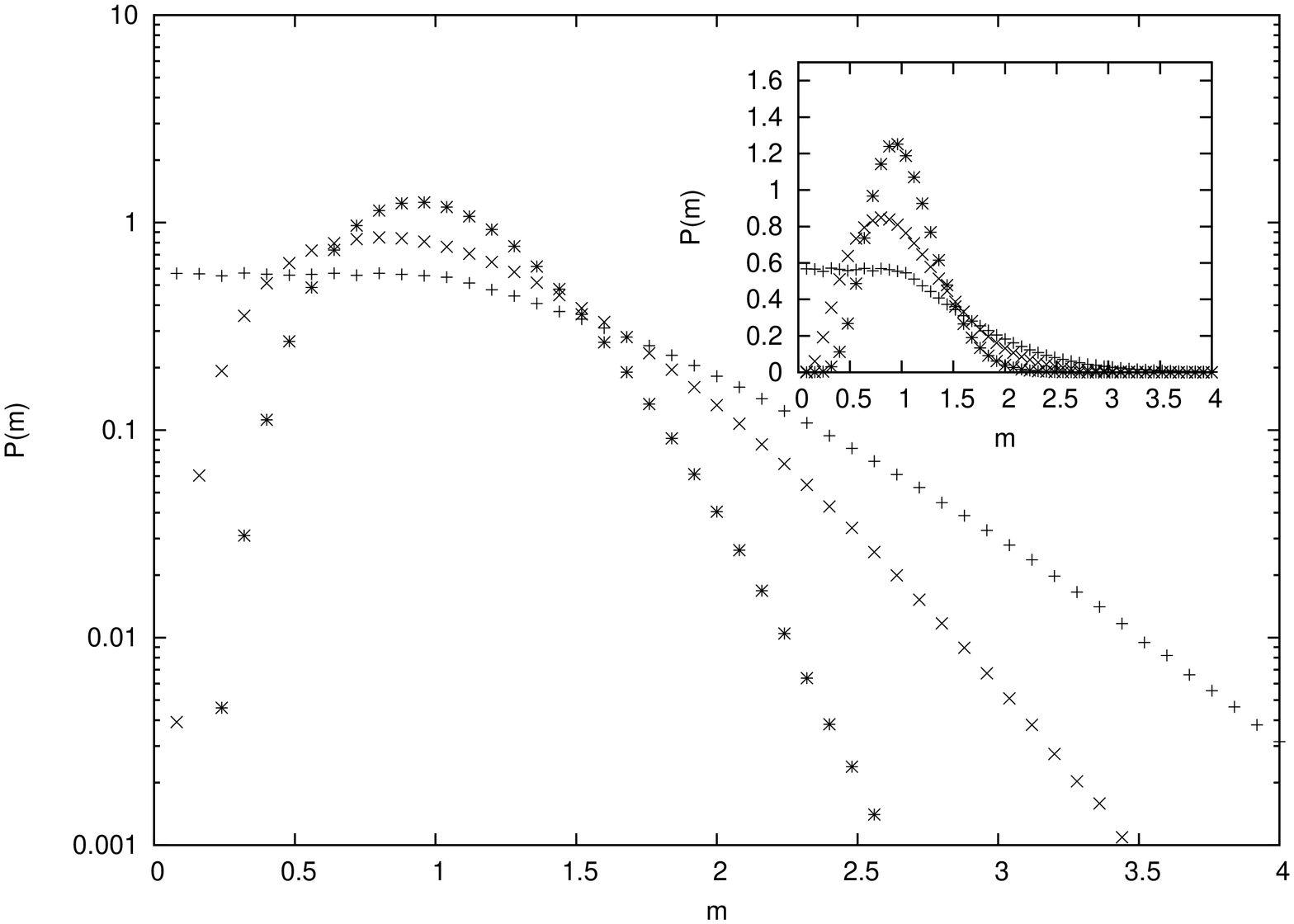}
\caption\protect{Steady state distributions generated by Eqn. \ref{gamma3} are plotted in the semi-log plot
: Three cases are shown
above, viz., $\lambda=0$ (+), $\lambda=0.4$ ($\times$),
$\lambda=0.7$ ($\ *$).  All simulations are done for $\sim ~10^5$ time steps and averaged over for $\sim ~10^4$.
{\it Inset} : The same data set in the usual plot.}
\label{almostgamma1}
\end{center}
\end{figure}

\noindent It is noteworthy that for $\lambda~=~0$, the distribution generated by Eqn. \ref{gamma3}
has a peculiar form which is known as Dickman distiburtion (see Ref. \cite{devroye;01}). 
For $m\leq1$, $P(m)$ is flat and for $m~>~1$, the distribution 
has a downward slope (see Fig. \ref{almostgamma1}). 
See Ref. \cite{chamayou;73}, for this type of maps which has been used in number theory,
in biology (see Ref. \cite{watterson;76}) and in many other areas (see Ref. \cite{knuth;76}). See
Ref. \cite{devroye;01} for simulations on the pdf generated by Eqn. \ref{gamma3} with $\lambda=0$.

It is to be noted that we cannot generate an exponential distribution from Eqn. \ref{gamma3} since its maximum variance is 1/2 (for $\lambda=0$) whereas the variance of the
exponential distribution generated by the following Eqn. (See Ref. \cite{acbkc;07})
\begin{equation}
m_i(t+1)=\epsilon\left(m_i(t)+m_j(t)\right)
\label{exp}
\end{equation}

\noindent is unity. Hence this model can not generate purely exponential distribution which is a drawback since
recent studies argue that the income/wealth distributions in the real world fits very well with exponential pdfs
\cite{yako-rosser;09, drag-yako;01, 
drag-yako;01a}. 
It is trivial to note that in the usual kinetic exchange models, it is the presence of the
trading partner's wealth $m_j(t)$ 
(in the $i$-th agent's wealth evolution equation; Eqn. \ref{exp}) that contributes to the higher variance.



\subsection{Power law distribution}
\label{subsec: Case 2}
\noindent 

\noindent We assume $\lambda_1 = \lambda$, $\lambda_2 =\lambda_3 = 1-\lambda$, 
$-\infty\le n \le 0 $ and $\theta \rightarrow \infty$.

\noindent The relevant equation is

\begin{equation}
m(t+1)= \{\lambda+\epsilon(1-\lambda)\}m(t)+\xi(1-\lambda)^n.
\label{power2}
\end{equation}

\noindent Ref. \cite{takayasu, sornette} studied this type of discrete stochastic equations (see Eqn. \ref{kesteneqn}
) as a generic model for
generating power law pdf. But a necessary condition for the mechanism to generate a power law pdf is that the random multiplicative term $a(t)$ ($\lambda+\epsilon(1-\lambda)$ in Eqn. \ref{power2}) must be greater than 1 sometimes.
In the current context, this condition is not satisfied (according to our interpretation, 
$(\lambda+\epsilon(1-\lambda))$ is the savings propensity and hence it is always less than unity).
To avoid this problem, we assume a population of agents each of which interacts with the market
according to Eqn. \ref{power2}. Note that the agents do not interact among themselves.
\noindent In the spirit of the kinetic exchange models, we show that if 
there is a population of $N$ agents with different $\lambda$ but 
each of the income evolution process is modelled by Eqn. \ref{power2}, then a power law in the
income distribution will be observed. However, there is an important difference with the usual kinetic
exchange models which are completely conservative. Any trading activity in such markets would be
a $zero-sum$ game i.e., if somebody wins then his/her trading partner has to lose. We relax that assumption here. 
For simulations, we assume that there are 200 agents each endowed with
an initial wealth equals to 1. Each of the agent's wealth evolution is governed by Eqn. \ref{power2}. Clearly,
there is no interactions between the agents. However, the agents are assigned different $\lambda$s which are
fixed over time for any given agent (see Ref. \cite{chatterjee;04}). In particular,  
we assume that $\lambda$ is uniformly distributed among
the agents. For each agent, simulations are done for $\sim ~10^4$ time-steps and the corresponding pdfs are 
averaged for $\sim ~10^3$ time-steps. Resulting distributions are averaged over all agents.

\begin{figure}
\begin{center}
\noindent \includegraphics[clip,width= 6cm, angle = 270]
{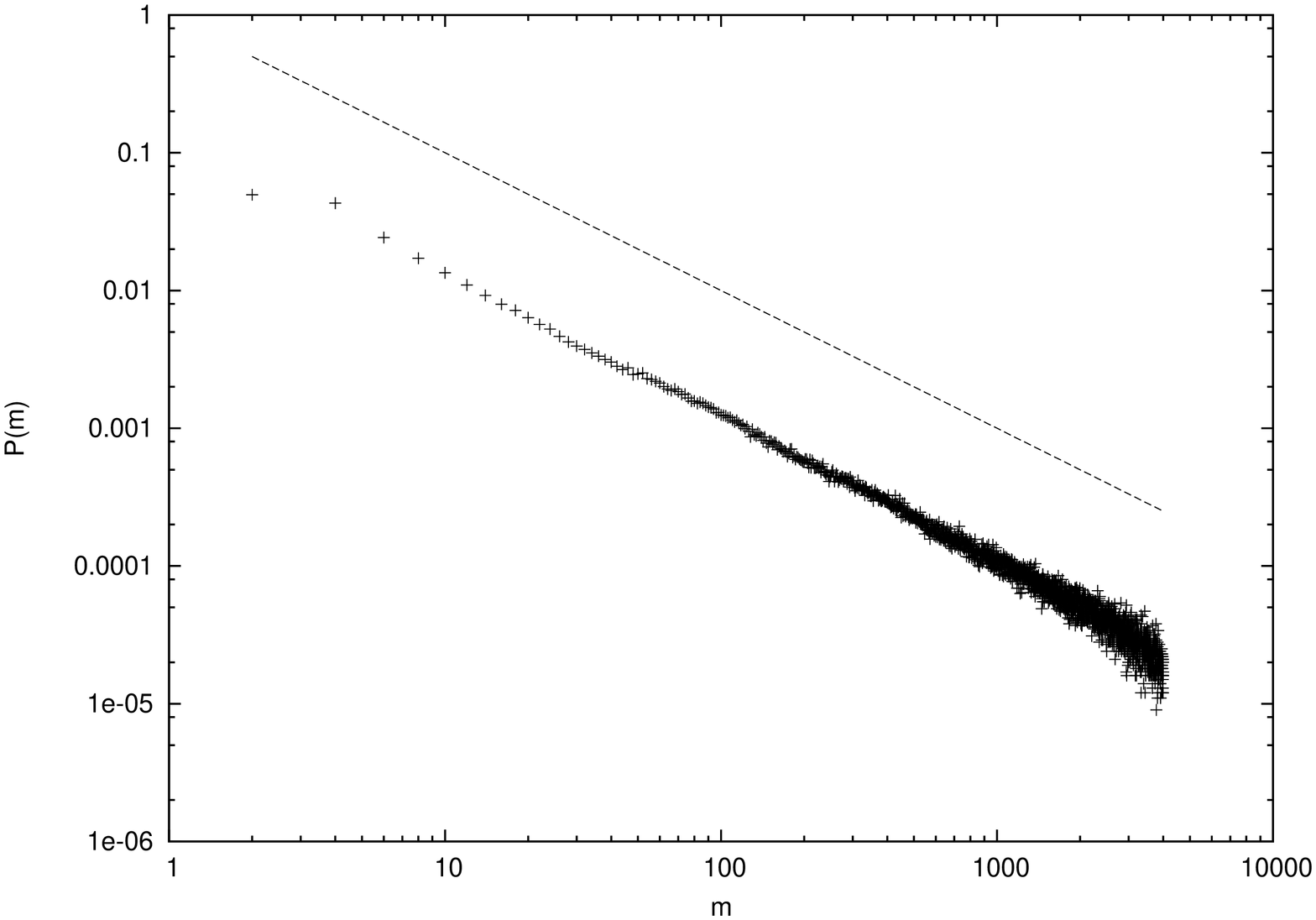}
\caption\protect{A power law distribution ($n$ = $-20$ in Eqn. \ref{power2}). 
The straight line with slope -1 is drawn as a guide. However, here we take 0$\le \lambda \le$0.33 because
otherwise the average value of $m$ becomes too large; e.g.,
Eqn. \ref{power3} shows that for $\lambda$ = 0.5, $\langle m\rangle$ = $2^{21}$.

}
\label{powerlaw1}
\end{center}
\end{figure}

\begin{figure}
\begin{center}
\noindent \includegraphics[clip,width= 6cm, angle = 270]
{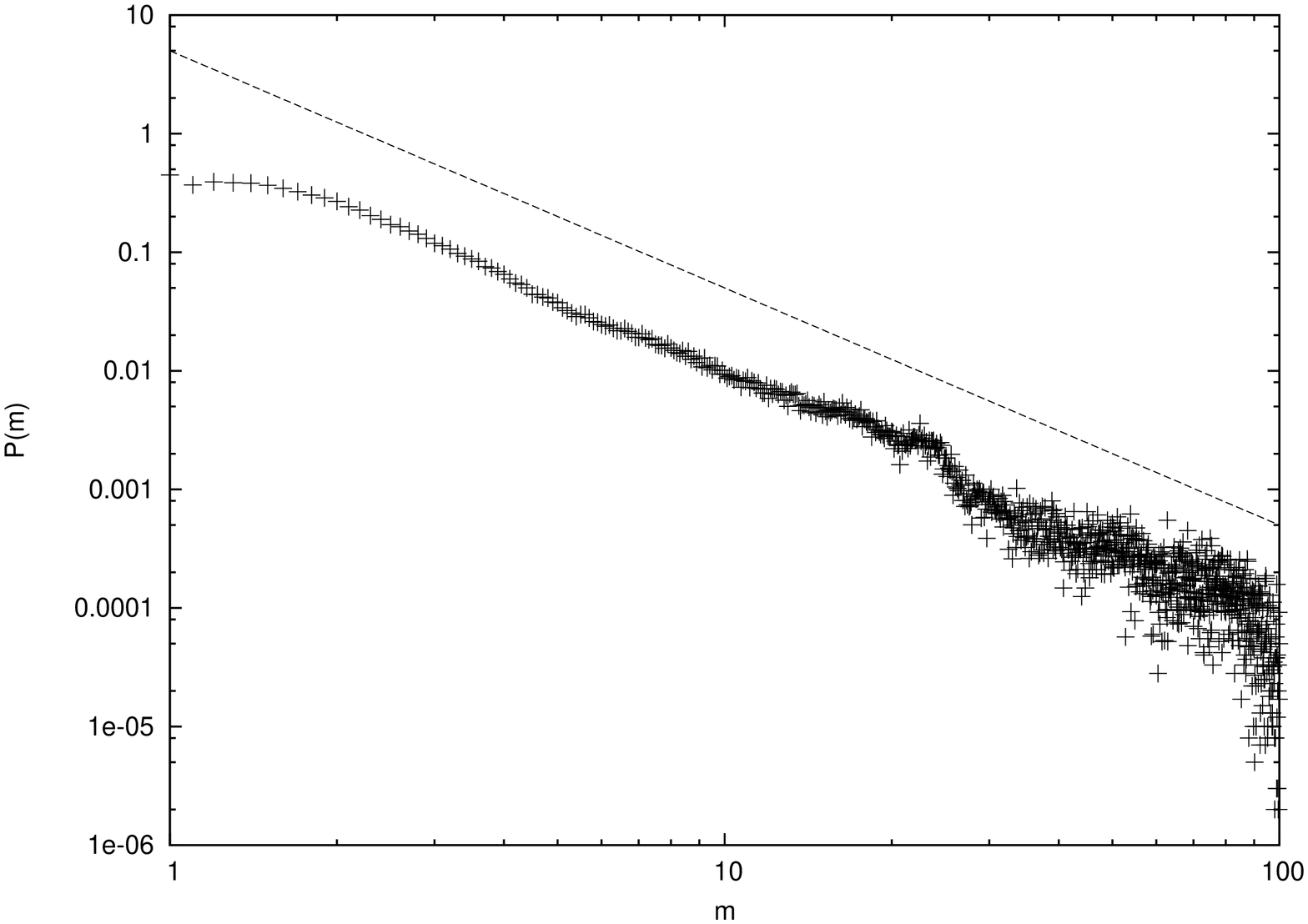}
\caption\protect{A power law distribution ($n$ = 0 in Eqn. \ref{power2}). The straight line with slope -2 is drawn as a guide.
}
\label{powerlaw2}
\end{center}
\end{figure}

\begin{figure}
\begin{center}
\noindent \includegraphics[clip,width= 6cm, angle = 270]
{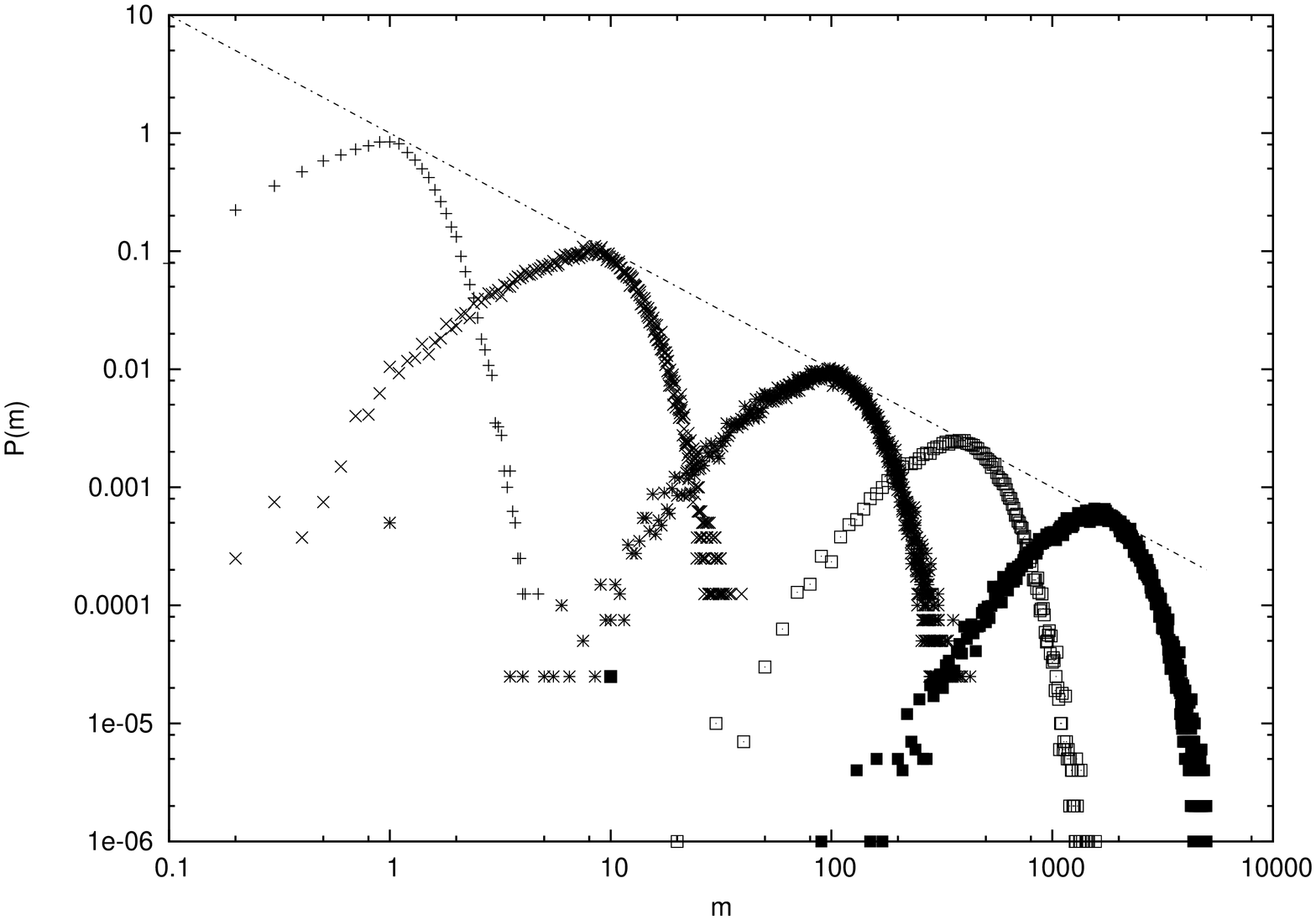}
\caption\protect{The pdf of wealth distributions are drawn for the case $n=-20$ with global 
savings propensities 0, 0.1, 0.2,
0.25, 0.3 (from the left to the right) in the log-log plot. Also, we've drawn $m^{-1}$ in the same diagram
(the dotted line). 
}
\label{powermode-20}
\end{center}
\end{figure}

\begin{figure}
\begin{center}
\noindent \includegraphics[clip,width= 6cm, angle = 270]
{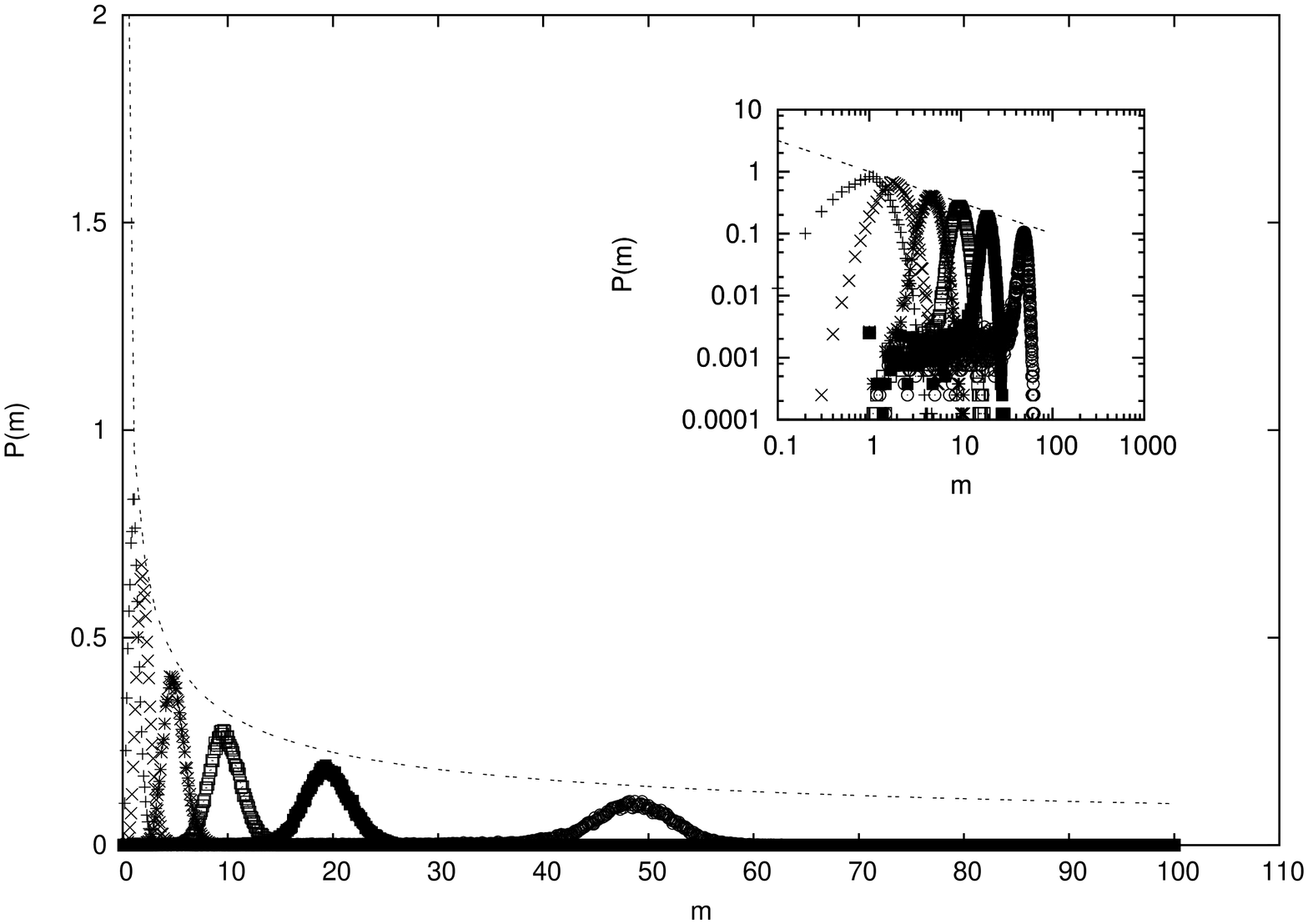}
\caption\protect{The pdf of wealth distributions are drawn for the case $n=0$ with global 
savings propensities 0.1, 0.5, 0.8,
0.9, 0.95, 0.98 (from the left to the right). Clearly, the average values of wealth at any given savings
propensity are very close to the theoretical values predicted by Eqn. \ref{power3}. 
Also, we've drawn $m^{-0.5}$ in the same diagram (the dotted line). 
{\it Inset}: The same diagram in log-log plot. }
\label{powermode0}
\end{center}
\end{figure}

\begin{figure}
\begin{center}
\noindent \includegraphics[clip,width= 6cm, angle = 270]
{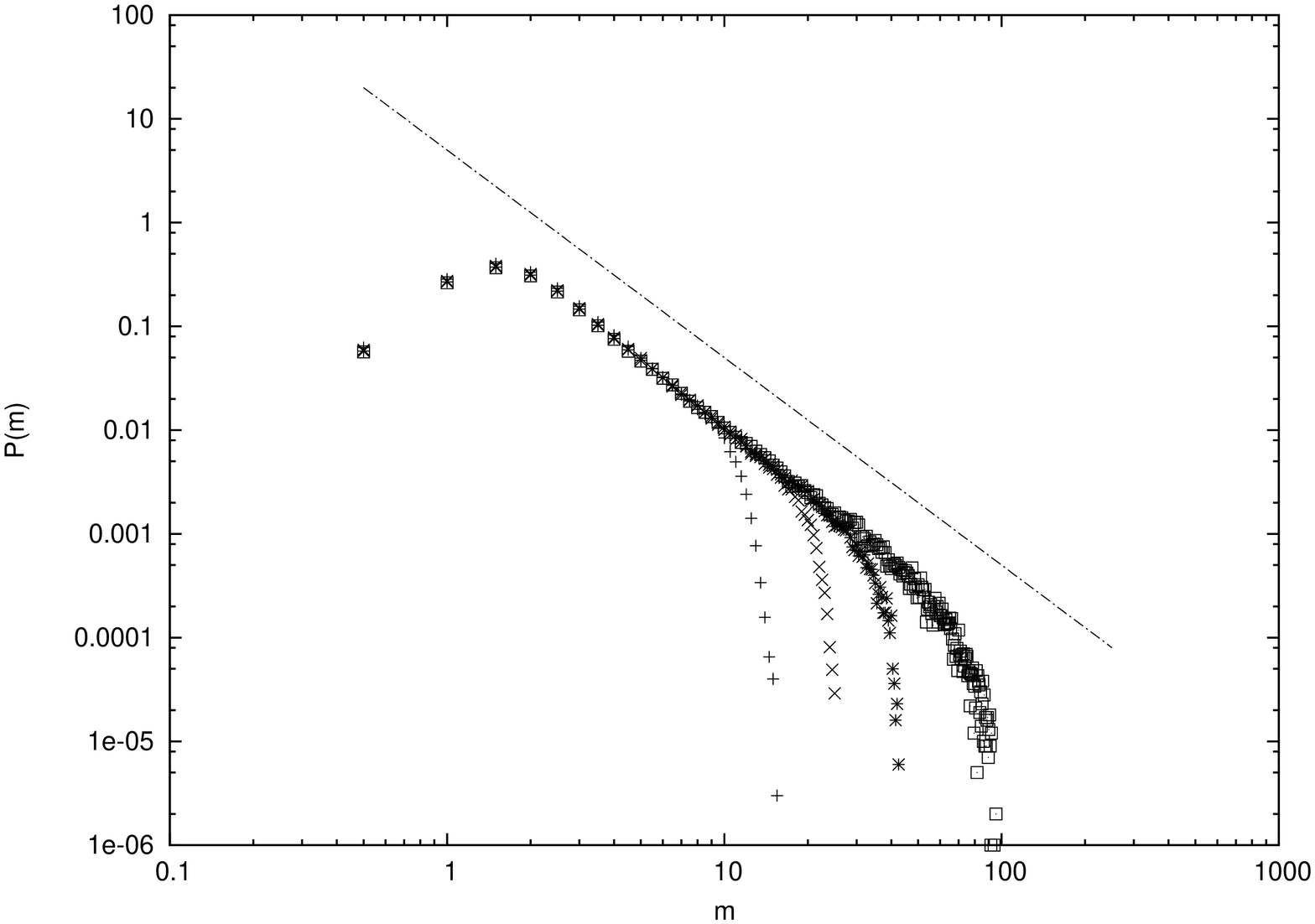}
\caption\protect{The pdf of wealth distributions are drawn for the case $n=0$ with 
maximum savings propensities $\lambda_M$ = 0.9, 0.95, 0.975,
0.99 (from the left to the right) in the log-log plot. The savings propensities of the agents
are chosen deterministically following the rule that $\lambda_i$ = $(i/N)\lambda_M$. 
Also, $m^{-2}$ is drawn for reference
(the dotted line). 
}
\label{powermaxl}
\end{center}
\end{figure}

\noindent Following Ref. \cite{mohanty}, a very simple proof is considered below.
Note that (in the steady state) by taking expectations on both sides of Eqn. \ref{power2}, we can rewrite it as
\begin{equation}
(1-\lambda)^{1-n}\langle m \rangle=1
\label{power3}
\end{equation}
\noindent By taking total differentiation and rearrenging terms, we get

\begin{equation}
\frac{d\lambda}{dm}=\frac{1}{1-n}m^{-\frac{n-2}{n-1}}
\label{power4}
\end{equation}

\noindent where $m$ represents $\langle m\rangle$. Hence, the average amount of money held by an agent with a 
particular $\lambda$ is given by Eqn. \ref{power3}. Also, the relation between the 
distribution of $\lambda$ (i.e., $\rho(\lambda)$) with that of $m$
is given by the following Eqn.

\begin{equation}
P(m) dm = \rho(\lambda) d\lambda.
\label{power5}
\end{equation}

\noindent Eqn. \ref{power4} and \ref{power5} shows that in an population of agents 
with uniformly distributed $\lambda$, the
distribution of $m$ would be

\begin{equation}
P(m)\sim m^{-\frac{n-2}{n-1}}.
\label{power6}
\end{equation}
 
\noindent It is trivial to note that the CCM model (See Ref. \cite{chatterjee;04}; See also Ref. \cite{acbkc;07}) 
derived the result analogous to the case where $n=0$ (see
Fig. \ref{powerlaw2}). Also, if $n$ is large in modulus then the steady state would be power law distribution
with power -1 (see Fig. \ref{powerlaw1}) (we take $n=-20$ just for expository purpose). 

\noindent Following Ref. \cite{acbkc;07}, we can support the above result for the case $n~=~0$. Let
us rewrite Eqn. \ref{power2} with the agent index as the following

\begin{equation}
m_i(t+1)= \{\lambda_i+\epsilon_i(1-\lambda_i)\}m_i(t)+\xi_i(1-\lambda_i)^n.
\label{power7}
\end{equation}

\noindent which can be rewritten (using Eqn. \ref{power3}) as

\begin{equation}
m_i(t+1)= \{\lambda_i+\epsilon_i(1-\lambda_i)\}m_i(t)+\xi_i(1-\lambda_i)\langle m_i\rangle.
\label{power8}
\end{equation}

\noindent The above equation holds true for all agents $i$. Since we focus on
average money holding at different savings propensities, we randomly pick any two agents $i$ and $j$ (with
savings propensities $\lambda_i$ and $\lambda_j$ respectively) and
multiply their corresponding wealth evolution equations (\ref{power8}), to get

\begin{eqnarray}
m_i(t+1)m_j(t+1)&=& f^1_{ij}m_i(t)m_j(t)+f^2_{ij}\langle m_i(t)\rangle m_j(t) \nonumber\\
&&+f^3_{ij}m_i(t)\langle m_j(t)\rangle + f^4_{ij}\langle m_i(t)\rangle \langle m_j(t)\rangle 
\label{power9}
\end{eqnarray}

\noindent where $f^k$s are functions of $\epsilon$, $\xi$, $\lambda_i$ and $\lambda_j$ for all $k$. We 
approximate each of the quadratic quantities by a mean quantity $m^2$. Note, that
here we assumed that $m_i$ can be replaced by its average value $\langle m_i\rangle$, which holds true
only if its variance is small and that requires $n$ to be as small in modulus as possible in Eqn. \ref{power7}. We 
consider $n~=~0$. Then, we have the following equation

\begin{equation}
m^2(t+1)= \alpha(t+1) m^2(t) ~~~~~ \mbox{where $\alpha$ is a function of $f^k$s. }.
\label{power10}
\end{equation}

Ref. \cite{pbacbkc;07,acbkc;07} studied this map and showed that as $t\rightarrow \infty$,
\begin{equation}
P(m)\sim m^{-2}.
\label{power11}
\end{equation}
\noindent Note that this argument holds only if $n=0$.

\noindent We consider the cases where all agents have the same savings propensities which we
call the global savings propensities (this case is almost identical to the one described in Section \ref{subsec: Case 1}
in that we consider agents who have the same $\lambda$; the difference is
that there we considered only one agent whereas here we consider $N$ identical agents. Also, another obvious difference 
is that here $n\le 0$). The pdfs are 
drawn for
individual savings propensities in Fig. \ref{powermode-20} and \ref{powermode0}. Let $m$ have a distribution
$f(m)$ in the steady state (given $\lambda$). 
With a slight abuse of notation, we write the value of $f(.)$ evaluated at $m=\bar{m}$ 
as $f(\bar{m})$. 
Then Fig. \ref{powermode-20} shows that (for $n=-20$) given $\lambda$ at $m=\langle m\rangle$,
$f_{\lambda}(\langle m\rangle)\approx  \langle m\rangle^{-1}$. Similarly, for $n=0$
given $\lambda$ at $m=\langle m\rangle$,
$f_{\lambda}(\langle m\rangle)\approx  \langle m\rangle^{-0.5}$. This shows that in the case where $n$
is large in modulus, the $f_{\lambda}(m)$ curve is a good approximation of the average of all the individual
wealth distributions which is not true in the other case where $n=0$. The reason is that
the probability of finding an agent within a specified interval of wealth decreases very rapidly with increase in wealth
in the case where $n=0$ compared to the other case (this can be checked very easily from Eqn. \ref{power3}).
Hence, though the uppermost points form a locus of $f_{\lambda}(\langle m\rangle)\approx  \langle m\rangle^{-0.5}$
for $n=0$, the average distribution falls much more rapidly following $P(m)\sim m^{-2}$.

\noindent We have also studied the effects of the maximum $\lambda$ present in the population on the steady state distribution (for the case $n=0$). Following Ref. \cite{germano;05}, we the assign the savings
propensities to the agents deterministically, according to the rule that 
the $i$-th agent's savings propensity $\lambda_i$ = $(i/N)\lambda_M$
where $N$ is the number of agents in the population, $\lambda_M<1$ is the maximum savings propensity
in the population. See Fig. \ref{powermaxl} for the results. Curves from left to right
represent the distributions for increasing $\lambda_M$. Clearly, as $\lambda_M$ decreases, the power law interval 
shrinks. Therefore, higher savings propensities contribute to the formation of the power law interval. This conclusion
is consistent with the findings in Ref. \cite{germano;05}. We should also mention that according to Eqn. \ref{power3},
higher $\lambda$ implies higher average wealth at that particular $\lambda$. This fact also 
supports the findings in Fig. \ref{powermaxl}.

\subsection{A non-trivial phase transition}
\label{subsec: Case 3}

\noindent Here, we briefly mention that a special case of the map shows a non-trivial phase transition. 
If we assume $\lambda_1=\lambda_2=\lambda$, $\lambda_3 = 0$, $n = 1 $ and $\theta = 1$.
The equation becomes

\begin{equation}
m(t+1)=min \{\lambda(1+\epsilon)m(t), 1\}
\label{phase1}
\end{equation}

\noindent This map has been used to model opinion formation. Assume that an agent can have an opinion $m(t)$  
within 0 and 1 at any time $t$. $\lambda$ is the agent's conviction parameter. After each interaction
with the society, his opinion changes by a random fraction. However, the maximum value of opinion that he
can have, is 1.  
Ref. \cite{lallouache;10} considered this map and showed that this map shows a phase transition
(with respect to $\lambda$)
in the average value of $m$. The critical value of $\lambda$ found from simulations, is $\lambda_c$ $\approx$ 0.68
and it only mentions an analytical result that $\lambda_c \approx$ 0.6796.
For numerical
details on this particular map, see Ref. \cite{lallouache;10}. Interestingly, this map has a parallel in the kinetic
exchange models, also discusses in the same reference (which has a critical point at $\lambda_c \approx 0.6667$). 
For detailed analysis on the nature of such transitions see Ref. \cite{sen;10,biswas;10}. 

\noindent Let us now focus on the cases with $\lambda<1$. Consider the map
\begin{equation}
m(t+1)=\lambda(1+\epsilon)m(t).
\label{phase2}
\end{equation}

\noindent Let us ignore $m(t)$ for all $t\le \tau$ where $\tau$ is sufficiently large. For any time step $T\ge\tau$,

\begin{equation}
m(T)=\lambda^{T-\tau}(1+\epsilon_{T})(1+\epsilon_{T-1})...(1+\epsilon_{T-\tau+1})m(\tau).
\label{phase3}
\end{equation}

\noindent Following Ref. \cite{lallouache;10}, we can argue that for $\lambda<\lambda_c$, 
the product of all the terms multiplied to $m(\tau)$ in the above equation is less than unity on average. But
this term tends to 1 as $\lambda\rightarrow\lambda_c$. 
Therefore at the critical point ($\lambda=\lambda_c$), by taking logs on both sides we can write
\begin{equation}
-\log~ \lambda_c=\frac{1}{T-\tau}\sum_k^T\log~(1+\epsilon_k).
\label{phase4}
\end{equation}
As $T\rightarrow\infty$, we apply the LLN and the r.h.s. of the above Eqn. converges to 
$\langle \log~ (1+\epsilon)\rangle$.
By applying Jensen's inequality, we see that
\begin{equation}
-\log ~\lambda_c < \log (\langle 1+\epsilon\rangle)
\label{phase5}
\end{equation}

\noindent i.e., $\lambda_c> 2/3$. Therefore the critical point of the map model is greater than 
that of its kinetic exchange version. By numerical calculations from Eqn. \ref{phase4}, we see
that $\lambda_c\approx 0.67954$ which is very close to the value reported in Ref. \cite{lallouache;10}. One could
also find that $\langle \log~ (1+\epsilon)\rangle$ = $2 \log 2-1$ and therefrom find $\lambda_c$ using Eqn. \ref{phase4}
(see \cite{lallouache;10}).

\noindent It is trivial to note that for $\lambda< \lambda_c$, $m(t)$ tends to 0. Also at $\lambda = \lambda_c$,
the distribution has an abrupt change in variance as has been 
shown numerically in Ref. \cite {lallouache;10}. Here, we present a
very short proof. By Eqn. \ref{phase3}, $\langle m(T)\rangle~=~\langle m(\tau)\rangle$ at $\lambda = \lambda_c$ and hence for $T-\tau$ sufficiently large,

\begin{equation}
m(T)-\langle m(T)\rangle\approx \lambda_c^{T-\tau}(1+\epsilon_T)(1+\epsilon_{T-1})...
(1+\epsilon_{T-\tau +1})(m(\tau)-\langle m(\tau)\rangle).
\label{phase6}
\end{equation}
\noindent By squaring and taking expectations on both sides, we get
\begin{eqnarray}
V(m(T))&\approx&\langle \lambda_c^{2(T-\tau)}(1+\epsilon_T)^2(1+\epsilon_{T-1})^2...(1+\epsilon_{T-\tau+1})^2\rangle V(m(\tau)) \nonumber\\
&>&  \langle \lambda_c^{T-\tau}(1+\epsilon_T)(1+\epsilon_{T-1})...(1+\epsilon_{T-\tau+1})\rangle ^2 V(m(\tau))\nonumber\\
&=& V(m(\tau)).
\label{phase7}
\end{eqnarray}

\noindent To get the last inequality, note that $V(x)=\langle x^2\rangle- \langle x\rangle^2 >0$ for all $x$
(unless it is the trivial case that $x=\langle x\rangle$). 
Therefore, $V(m(T))>z_{(T-\tau)}V(m(\tau))$ with $z_k > 1$ for all $k$ and for all $\tau$.
Hence, as $t\rightarrow\infty$, $V(m(t))\rightarrow \infty$ at $\lambda=\lambda_c$.

\noindent To illustrate the problem of non-stationarity, we focus on another particular case where $\lambda\rightarrow 1$,
$\epsilon \sim uniform[0,\bar{\epsilon}]$ with $\bar{\epsilon}\rightarrow 0$ such that 
$\langle \epsilon\rangle = - \log~\lambda$ i.e., $\langle \epsilon\rangle = - \log~(1-(1-\lambda))\approx 1-\lambda$.
From Eqn. \ref{phase3}, by taking logs we get
\begin{eqnarray}
\log~ m(T)&=& (T-\tau)\log~ \lambda+\sum _{k}^T \log ~(1+\epsilon_k)+log ~m(\tau)\nonumber\\
&\approx&\sum_{k}^T (\epsilon_k-\langle \epsilon_k\rangle)+\log ~m(\tau).
\end{eqnarray}

\noindent Multiplying both sides by $1/\sqrt{T}$ and by applying CLT (assuming $T\rightarrow \infty$), we get

\begin{equation}
\frac{1}{\sqrt{T}}\log~ m(T) ~\rightarrow~ N(0,\sigma^2) ~~~~~\mbox{in distribution,}
\label{lognormal0}
\end{equation}
\noindent where $\sigma^2$ is the variance of $\epsilon$. Hence, $m(T)$ is distributed log-normally. But evidently, the distribution is not stationary.

\subsection{The law of proportionate effect}
\label{subsec: Case 4}
\noindent 

\noindent We assume $\lambda_1 = \lambda_2= 1$, $\lambda_3 = 0$ and $\theta \rightarrow \infty$. Note that this
is the limit of the case discussed above with $\lambda = 1$ and $\theta$ sufficiently large. In this case,
the equation becomes

\begin{equation}
m(t+1)=(1+\epsilon_{t+1})m(t)
\label{lognormal}
\end{equation}

\noindent which is very wellknown as a generator of log-normal distribution (See Ref. \cite{gibrat;31}). 
To find the distribution of $m(t)$, note 
that the above equation can be rewritten (by iteration) as

\begin{equation}
m(t+1)= \prod_k^t(1+\epsilon_k)m(1)
\end{equation}

\noindent and by taking log on both sides, it can be written as

\begin{eqnarray}
\log~ m(t+1)&=& \sum _{k}^t \log ~(1+\epsilon_k)+log ~m(1)\nonumber\\
&\approx &\sum_{k}^t \epsilon_k+\log ~m(1) ~~~~~~~\mbox{(for $\epsilon$ very small).}\nonumber\\
\end{eqnarray}

\noindent By applying the Central Limit Theorem, we see that $\log ~m(t)$ is distributed normally, hence $m(t)$
is distributed log-normally. However, evidently the distribution generated this way is not stationary either.

\section{Summary}
{
\noindent In recent years, a number of economic regularities have been investigated in the econophysics
literature, one of the most prominent themes being the distributions of income/wealth.
The two candidate distributions for explaining the left tail and the mode of the distributions of income/wealth
are Gamma and Log-normal. There is a consensus that the heavy right tail is best described by a power 
law (See Ref. \cite{yako-rosser;09,acbkc;07}).
Also, the process of opinion formation due to interactions among numerous agents have been studied in
details. The phenomena of emerging consensus and a phase transition in the opinion formation
have been tried to be modelled in many ways (See Ref. \cite{weisbuch,lallouache;10}).

\noindent In this paper, we examine a stochastic map to model the economic process of income/wealth
distribution and the sociological process of opinion formation. We showed that this
particular map can generate a positively skewed pdf and a power law pdf. It
shows a non-trivial phase-transition and in another limit, 
it coincides with a very well known generator of log-normal pdf.
However, there are several drawbacks of this approach. One is that the algebraic form of the pdf generated by
the Eqn. \ref{gamma1} is not known and it is not a $\Gamma$ pdf (at one limit it has a sharp peak which 
eventually goes away). 
Second is the well known
problem with Eqn. \ref{lognormal} that the distribution generated by this equation is not stationary. Third,
purely exponential distribution cannot be generated by this model. But there are a number of recent studies concluding
that the income/wealth distribution can be modelled by exponential pdf \cite{yako-rosser;09, drag-yako;01, drag-yako;01a}.
Previously, the kinetic exchange models have been successful to model many of these phenomena 
\cite{acbkc;07,lallouache;10}. 
We show that there
are some new results produced by this map model whereas other results mostly
conform with those derived in the kinetic exchange models.
 
}

\medskip

{\bf Acknowledgement}

\medskip

{
\noindent I am grateful to Arnab Chatterje and Bikas K. Chakrabarti for some useful discussions
and careful reading of the manuscript.
}

\end{document}